\documentclass[journal]{IEEEtran}
\usepackage{hyperref}
\usepackage[cmex10]{amsmath}
\usepackage{amsfonts}
\usepackage{amssymb}
\usepackage{pgf}
\usepackage{graphicx}
\usepackage{grffile}
\usepackage{subfig}
\usepackage{epstopdf}
\usepackage{mathtools}
\usepackage{mathabx} % for \not\perp

\ifCLASSINFOpdf
\else
\fi
\begin{document}
\onecolumn
\title{Adaptive Mean Queue Size and Its Rate of Change: Queue Management with Random Dropping}

\author{Karmeshu$^\star$, Sanjeev Patel$^\star$, and Shalabh Bhatnagar$^\dagger$\thanks{$^\star$School 
of Computer and Systems Sciences, Jawaharlal Nehru University, New Delhi - 110067.}\thanks{$^\dagger$Department of 
Computer Science and Automation, Indian Institute of Science, Bangalore 560012.}}

\maketitle


\begin{abstract}
The Random early detection (RED) active queue management (AQM) scheme uses the average queue size to calculate the dropping 
probability in terms of minimum and maximum thresholds. The effect of heavy load enhances the frequency of crossing the maximum 
threshold value resulting in frequent dropping of the packets. An adaptive queue management with random dropping (AQMRD) algorithm 
is proposed which incorporates information not just about the average queue size but also the rate of change of the same. 
Introducing an adaptively changing threshold level that falls in between lower and upper thresholds, our algorithm demonstrates 
that these additional features significantly improve the system performance in terms of throughput, average queue size, 
utilization and queuing delay in relation to the existing AQM algorithms.
\end{abstract}


% \fi
\IEEEpeerreviewmaketitle

\section{Introduction}
\IEEEPARstart{A}{ctive queue management} is the most effective network assisted algorithm to control the congestion at the routers. There are several AQM algorithms proposed in the literature viz. drop-tail, random early detection (RED), random early marking (REM) etc. RED has been able to enhance the throughput by using dropping function in terms of the average queue size~\cite{sp1}. A desirable feature of the algorithm is to reduce the loss rate in the presence of random traffic characteristics. Another algorithm REM introduces rate mismatch as well as queue mismatch to yield high throughput or achieve low loss-rate~\cite{sp3}. Queue mismatch is the difference between target queue length and current queue length and rate mismatch is the difference between link capacity and input rate. REM differs from RED since it uses a different measure for congestion and hence has a different technique for calculating the marking / dropping probability. This congestion measure or price is updated based on both mismatch observed for queue and rate. Similarly, REM computes the prices for each link and then calculates their sum to determine the end-to-end marking probability. The latter probability increases with higher congestion measures or link prices~\cite{sp3}. A new algorithm BLUE improves the performance by reducing the loss rate of the packets by modifying the dropping function based on loss events~\cite{sp4}. A separate first-come-first-served (FCFS) queue is required in fairness-queuing (FQ) scheme for each conversation. The queues are serviced in a round-robin fashion so as to allocate equal bandwidth to each queue. Stochastic fairness queuing (SFQ) is proposed to avoid infeasible computational requirements for high-speed networks~\cite{sp5}. In the case of large number of queues as compared to the number of conversations, a high probability is provided for each conversation being assigned to its own queue. The flow or pair of source and destination receives less than the allocated shared bandwidth when two conversations collide. SFQ provides a mechanism so that collided conversations for one slot are very unlikely to collide during the next~\cite{sp5}.

  A variant of RED has been proposed in~\cite{sp6} to resolve the limitations observed with traditional RED~\cite{sp2}. Refining RED, Wang et. al.~\cite{sp7} have enhanced the throughput by keeping the average queue size below the threshold value to avoid dropping of the packets. This has been realized by first adapting the queue weight parameter and thereafter the values of the $max_p$ parameter are chosen to stabilize the queue size. A limitation with RED is that packets are discarded even when the queue size is lower than the threshold value. Feng et. al.~\cite{sp8} have addressed the issue of unnecessary dropping of packets by requiring additional information on instantaneous 
  queue size. The important aspects for the stabilization of queue in the RED gateway are discussed in~\cite{sp11}-\cite{sp13}. A robust optimization technique for RED is analyzed in~\cite{sp14}. This robust optimization technique is independent of technology, model and protocols used. A detailed comparative study of AQM algorithms can be found in~\cite{sp15}. In RED, the queue size varies according to congestion level which leads to unpredictable queuing delay. A major issue with traditional RED has been the setting of the parameters and their tuning in order to achieve good performance~\cite{sp2}. Adaptive RED achieves significant control resulting in improved throughput using dynamic adaptation of the maximum probability $(max_p)$ parameter. Feng et al.~\cite{sp21} have proposed a three-section RED (TRED) to overcome this issue particularly for heavy loads by dividing the queue size interval $(max_{th}-min_{th})$ into three equal sections. In TRED, dropping probability is calculated according to the dropping function used in the three different sections. Nonlinear dropping functions are used in lower and upper sections of the queue size interval. In addition, the middle interval uses a linear dropping function. TRED is able to improve the throughput at low-load and maintains low delays at high-load~\cite{sp21}. 
  
  The problem of parameter setting in TRED is not adddressed as in RED. Further, Adaptive RED also does not show good performance under high load conditions. Also, in the case of proportional controller, it is found to suffer from a limitation under certain situations in which it is infeasible to implement. The instability is found in proportional controller if an operating point of $p$ lies between $p_{max}$ and 1 for such network conditions which result in oscillations of queue length. This leads to increase in the queuing delay and this issue arises due to the coupling between the average queue size and the dropping probability. The Proportional Integrator (PI) controller has been proposed to decouple both the average queue size and the dropping probability~\cite{sp9}. The PI controller is implemented with the nonlinear TCP dynamic by introducing the role of the queue's operating point. Wu et al.~\cite{sp16} have proposed an interactive mobile streaming scheme which has the capability to maintain a certain level of service quality in the presence of dynamic network environments. Most of the AQM schemes have been studied for wired networks. Adaptive Optimized Proportional Controller (AOPC) is another novel AQM scheme based on controlling the queue parameter~\cite{sp18}. The dropping probability is updated for a very small time interval upon each packet arrival in AOPC. AOPC measures packet loss ratio for larger intervals by introducing load estimator for the network. TCP load is estimated 
with the help of packet loss ratio and then TCP/AOPC feedback is optimized according to the second-order system model~\cite{sp18}. It makes AOPC stabilize the queue length very close to the target queue length using an optimized second-order system model. Chavan et. al.~\cite{sp19} have, however, discussed the robust AQM for wireless networks where a challenging problem results from fading and the resulting change in bandwidth allocation~\cite{sp14}. Chavan et. al.~\cite{sp16} have designed a better queue controller by tuning the operating points offline. An analytical model for bursty and correlated traffic to compute performance measures such as end-to-end delay, loss probability, and throughput is also proposed in ~\cite{sp20} in the presence of hybrid wireless networks. This analytical model is also applicable for bursty and correlated traffic and has been tested on OMNeT++ simulator.
 
One of the serious limitations of the aformentioned algorithms is that they are based on information regarding the mean queue size alone. 
However, in reality the rate of change of queue size on account of non-stationary heavy traffic may provide much deeper insights into the 
growth dynamics of queue build up. Our work aims to incorporate information both about the mean queue size at any time and its rate of change.
In fact, prior work completely ignores this second-order dynamics that our work aims to capture. Our work makes our system akin to
a dynamical system where not just the velocity but also information on acceleration is captured through the system measurements
and this new information is also used to control the system dynamics. We observe that this (latter) information indeed helps obtain better
control over the system.

%The existing algorithms of AQM do not consider the rate of change of queue size which may also play a vital role in the increase of %the number of received packets. This aspect assumes considerable significance when one deals with bursty traffic. Introducing %additional threshold value, viz, mid-threshold ($mid_{th}$) lying between  $min_{th}$ and  $max_{th}$, a generalized approach is %proposed to analyze the throughput, mean queue size, utilization, and queuing delay.

This paper comprises of five sections. Section II discusses a queue rate based model which is introduced to overcome the problems observed in the existing algorithms. The simulations and discussions are presented in section III. Section IV evaluates the performance of our proposed scheme with existing schemes using ns-2 simulator. In the last section, we provide the concluding remarks. 
     
\section{System Model}
\subsection{Queue-Rate Based Model}
A new approach which incorporates both queue size and its rate of change would better characterize the evolutionary dynamics of queue size build up. The dropping function is a key variable which affects the throughput or the loss-rate in the AQM algorithm. The dropping function in RED is based on the assumption that the input traffic characteristics do not  change much with time. In contrast to RED, we have introduced a new parameter $davg$, corresponding to the rate of change of the average queue size in addition to the average queue size ($avg$) itself. The proposed AQMRD approach calculates the quantities $avg$ and $davg$ according to 
\begin{equation}
 avg(t+1) = (1-w_q)avg(t)+w_qq(t),
\end{equation}
\begin{equation}
 davg(t+1) = (1-w_q)davg(t)+w_q ( q(t) - q (t-1)),
\end{equation}
where $avg(t)$ and $davg(t)$ are the average queue size and the rate of change of the average queue size respectively at time $t$. We increment time by one unit so that sufficient number of packets arrive/depart during the interval. Accordingly, the information about rate of change of the average queue size would reflect the traffic characteristics to achieve better congestion control at the routers. A positive value of the rate of change of $avg$ indicates that the queue size attains the threshold rapidly whereas a negative rate of change indicates a slow-down in the process of reaching the threshold. 

In order to prevent packet drop at the gateways, our aim is to modify the dropping function so that unused buffer space remains available to accommodate the new arriving packets. This is achieved by noting that a positive value of $davg$ results in a reduction in the number of received packets. RED has been found to perform well in the case of moderate traffic. One of the deficiencies of RED is that it is hard to configure and is sensitive to traffic load. Our proposed algorithm is able to capture the changing nature of non-stationary traffic. In the RED gateway, the average queue size is compared to two thresholds, maximum threshold and minimum threshold. However, the effect of non-stationary heavy traffic results in the increase in frequency of crossing or reaching the $max_{th}$  and thus resulting in more packets being dropped. In order to address this limitation, we have modified the RED algorithm by introducing a new variable, viz., $mid_{th}$ lying between $min_{th}$ and $max_{th}$, which adapts to rapidly changing temporal variation of $avg$. The proposed extended framework results in qualitatively much improved values of the performance metrics. To avoid reduction in the number of received packets, we have applied a more aggressive dropping function by varying $mid_{th}$ towards $min_{th}$.

This allows significant reduction in the number of discarded packets and is achieved when a gateway marks the packet more aggressively before $avg$ reaches the  $max_{th}$. As a negative rate of change will slow down the process of reaching   $max_{th}$, it will not adversely affect the dropping of the packets. However, a positive rate of change is likely to force the system to enter into an undesirable state resulting in dropping of several packets. The introduction of a   $mid_{th}$ takes care of such an adverse situation.  The proposed queue-rate based model allows more unused space to be left in the buffer, thus reducing the number of times  $avg$ crosses the maximum threshold. Consequently, it results in an increase in the number of received packets by dynamically updating $mid_{th}$ depending on the value of $davg$.

\subsection{Proposed Algorithm: AQMRD}
In the light of the significant role played by $davg$, we have proposed a new queue-rate based algorithm for the gateways. The AQMRD gateway performs different decisions for $mid_{th}$ according to (3). Here, $mid_{th}$ is updated as follows: 
\begin{equation}
mid_{th}=\begin{cases}
mid_{th}+1& \text{if $davg<0$},\\
mid_{th}-1& \text{if $davg >0$},\\
mid_{th}& \text{if $davg=0$}.
\end{cases}
\end{equation}

The dropping probability function is calculated as follows:

\begin{equation}
p_b=\begin{cases}
p_1& \text{if $davg > 0$},\\ \tag{4.a} 
p_2& \text{if $davg \leq 0$},\\
\end{cases}
\end{equation}
where $p_1$ and $p_2$ are given by
\begin{equation}
p_1=\frac{avg-min_{th}}{mid_{th}-min_{th}}max_p, \tag{4.b} 
\end{equation}
\begin{equation}
p_2=\frac{avg-min_{th}}{max_{th}-min_{th}}max_p. \tag{4.c} 
\end{equation}
For a positive value of $davg$, the AQMRD gateway starts dropping packets when $avg$ crosses the threshold $mid_{th}$ rather than $max_{th}$. The AQMRD gateway for positive $davg$ calculates the dropping probability according to
\begin{equation}
p_b=\begin{cases}
0& \text{if $min_{th}>avg$},\\
\frac{avg-min_{th}}{mid_{th}-min_{th}}max_p& \text{if $min_{th}\leq avg<mid_{th}$},\\ \tag{5}
1& \text{if $mid_{th} \leq avg$},
\end{cases}
\end{equation}
where $mid_{th}$ is a variable which varies between $min_{th}$ and $max_{th}$ according to (3). In the case of a non-positive value of $davg$, the dropping probability is computed as 
\begin{equation}
p_b=\begin{cases}
0& \text{if $min_{th}>avg$},\\
\frac{avg-min_{th}}{max_{th}-min_{th}}max_p& \text{if $min_{th}\leq avg<max_{th}$},\\ \tag{6}
1& \text{if $max_{th} \leq avg$}.
\end{cases}
\end{equation}\\ \\

\begin{figure}[!h]
\centering
\includegraphics[angle=0,width=0.58\textwidth,height=0.36\textwidth]{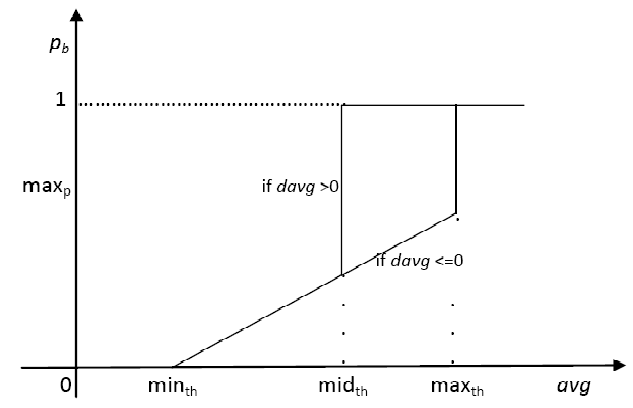}
% where an .eps filename suffix will be assumed under latex,
% and a .pdf suffix will be assumed for pdflatex; or what has been declared
% via \DeclareGraphicsExtensions.
\caption{AQMRD's packet dropping function}
\label{fig_dropfunct}
\end{figure}

\textbf {The Proposed AQMRD Algorithm:}\\ 
\ttfamily
\\Initialization: \\
   \hspace*{1.3 cm} $count \leftarrow -1$\\
for each packet arrival\\
 \hspace*{0.5 cm}calculate the average queue size $avg$ and rate of change of average queue size $davg$\\
  \hspace*{1.3 cm}if $min_{th} \leq avg < max_{th}$\\
   \hspace*{1.8 cm} increment $count$\\
   \hspace*{1.8 cm} if $davg > 0$\\
   \hspace*{2.5 cm} decrement the $mid_{th}$\\
   \hspace*{2.5 cm} if $avg<mid_{th}$\\
      \hspace*{3 cm}calculate probability $p_1$\\
      \hspace*{2.5 cm} else \\
    \hspace*{3 cm}calculate probability $p_2$ \\
    \hspace*{2.5 cm}calculate probability $p_b$ using $p_1$ and $p_2$\\
      \hspace*{2.5 cm}update the dropping probability \\
          \[p_a \leftarrow \frac{p_b}{1-count.p_b}\]
      \hspace*{2.5 cm}mark the arriving packet with probability $p_a$\\
   \hspace*{2 cm}else \\
    \hspace*{2.5 cm} increment the $mid_{th}$\\
      \hspace*{2.5 cm}calculate probability $p_1$\\
      \hspace*{2.5 cm} calculate probability $p_b$ using $p_1$\\
      \hspace*{2.5 cm} update the dropping probability \\
          \[p_a \leftarrow \frac{p_b}{1-count.p_b}\]
      \hspace*{2.5 cm}mark the arriving packet with probability $p_a$\\
   \hspace*{2 cm} if gateway mark the arriving packet\\
     \hspace*{2.5 cm} $count \leftarrow 0$\\
    \hspace*{0.8 cm} else if $max_{th} \leq avg$\\
       \hspace*{1.2 cm} $count \leftarrow -1$\\

\rmfamily

The $min_{th}$ is not kept too low so that the AQMRD gateway does not underutilize the bandwidth, particularly in view of fluctuations arising on account of heavy and non-stationary traffic. Depending upon the permissible delay, the threshold $max_{th}$ is set. If the difference between $max_{th}$ and $min_{th}$  is small then average queue size is likely to reach the maximum queue size frequently as seen in the drop-tail scheme. Accordingly, to circumvent the problem, we choose $mid_{th}$ so that it varies according to
   \begin{equation}
    mid_{th}=x. min_{th}   \text{  where $x\in\lbrack1,3\rbrack $}. \tag{7}
\end{equation}  
In case of a positive value of $davg$, the difference between $mid_{th}$ and $min_{th}$ should be sufficient enough to avoid global synchronization. AQMRD's dropping function is shown in Fig.~\ref{fig_dropfunct}. The advantage of the proposed approach is that it does not suffer from the phenomenon of global synchronization due to the dynamic nature of $mid_{th}$. The advantage of AQMRD is that it achieves a better throughput for heavy traffic with the help of prior information regarding the change in average queue size and the rate of change of the same. This dynamic adaptation of $mid_{th}$ results in an adaptive nature of queue management and which in turn results in improvement of the performance irrespective of the parameter setting. The next section presents a comparative study of our scheme with existing AQM schemes using the ns-2 network simulator.

%\textbf{Parameter setting of $mid_{th}$}

\section{Simulations and Discussions}
\subsection{Simulation Setup} 
%We consider heavy input traffic with power law characteristics. To this end, we follow the work of Feldmann and Whitt who use the %theorem due to Bernstein. For the sake of completeness, we reproduce the theorem~\cite{sp10}.\\

%\textbf{Theorem.} \textit{Every completely monotone pdf f is a mixture of exponential pdf 's, i.e.,}
%\begin{equation}
%f(t) =\int_0^\infty \lambda e^{- \lambda t} dG(\lambda), \text{ $t \geq 0$}, \tag{8}
%\end{equation} 
%\textit{for some proper cdf of G.}\\ 

%For simulating power law characteristics, we employ a finite mixture of exponentials which approximates a long-tail distribution. %Superimposing outputs of several independent sources, with each source following exponential inter-arrival time distribution, we %observe a significant effect upon performance of the network. To achieve this, we have taken 
We consider the network topology as shown in Fig. 2. The gateway/router parameters used in the experiments are set as $w_q=0.002$, $max_p=0.1$, and $max_{th}=48$ packets. The queue parameter $min_{th}$ is set at one-third of $max_{th}$ in each simulation to achieve good performance~\cite{sp5}. A gateway has a buffer size of 64 packets and each FTP source sends a packet with a maximum packet size of 1000-bytes until the congestion control window allows sending of the packets.
\begin{figure}[!h]
\centering
\includegraphics[angle=0,width=0.68\textwidth,height=0.42 \textwidth]{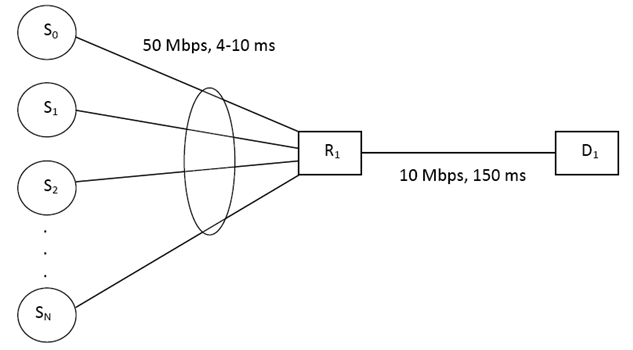}
% where an .eps filename suffix will be assumed under latex,
% and a .pdf suffix will be assumed for pdflatex; or what has been declared
% via \DeclareGraphicsExtensions.
\caption{Network topology}
\label{netscenario}
\end{figure}
For performing simulation, we assume random propagation delay varying from 4 ms to 10 ms between the FTP sources and router. Each simulation is run for a duration of 100 seconds. We choose maximum probability $max_p=0.1$ to check the efficacy of the proposed scheme with respect to the parameter setting. Due to the adaptive nature of our proposed scheme as discussed later, AQMRD achieves good performance even for a high maximum probability as $max_p=0.1$, indicating that our scheme has low sensitivity towards the $max_p$ parameter because at a low value of $max_p$, our scheme achieves good performance. We set the delay bandwidth product at around 200 packets which equals the congestion window size~\cite{sp5}. 

\subsection{AQMRD's Parameter Settings}
\begin{figure}[!h]
\centering
\includegraphics[angle=0,width=0.6\textwidth,height=0.5\textwidth]{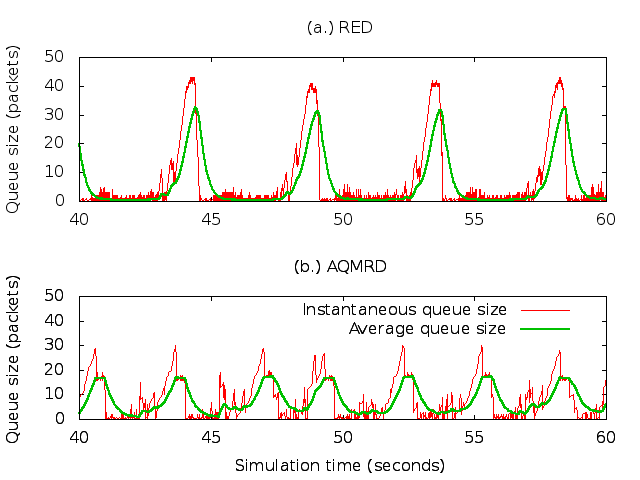}
% where an .eps filename suffix will be assumed under latex,
% and a .pdf suffix will be assumed for pdflatex; or what has been declared
% via \DeclareGraphicsExtensions.
\caption{AQMRD vs RED: comparative changes in $q$ and $avg$ for $N=25$}
\label{fig_queue25}
\end{figure}

In our experiments, we have simulated settings with 25, 50, 75 and 100 FTP sources, respectively. These correspond to low, moderate, high and very high traffic load conditions. For our first set of experiments, we have simulated 25 FTP sources for the network topology shown in Fig.~\ref{netscenario}. We show in Fig.~\ref{fig_queue25} the behavior of both average and instantaneous queue sizes in RED and AQMRD gateways for these 25 FTP sources. This shows that the AQMRD gateway suppresses the fluctuations in the average queue size and instantaneous queue size by more than what is observed in the RED gateway. This lower value of average queue size reduces the queuing delay in our scheme.

\begin{figure}[!h]
\centering
\includegraphics[angle=0,width=0.6\textwidth,height=0.5\textwidth]{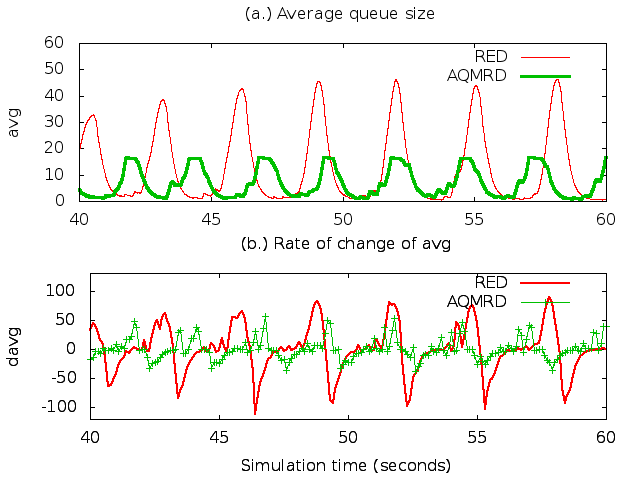}
% where an .eps filename suffix will be assumed under latex,
% and a .pdf suffix will be assumed for pdflatex; or what has been declared
% via \DeclareGraphicsExtensions.
\caption{AQMRD vs RED: comparative changes in $avg$ and $davg$ for $N=50$}
\label{arate}
\end{figure}

\begin{figure}[!h]
\centering
\includegraphics[angle=0,width=0.6\textwidth,height=0.5\textwidth]{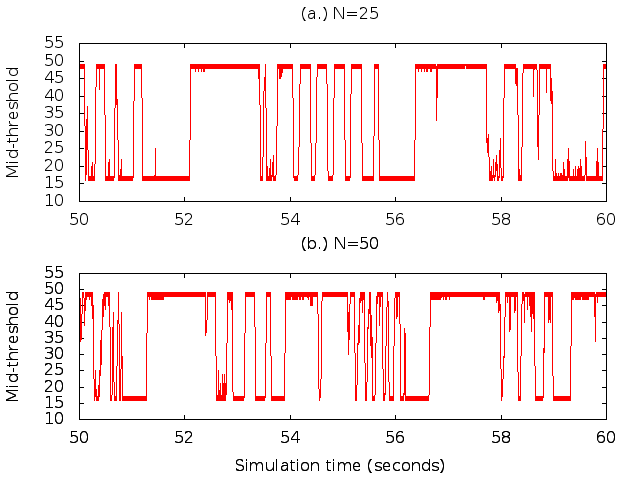}
% where an .eps filename suffix will be assumed under latex,
% and a .pdf suffix will be assumed for pdflatex; or what has been declared
% via \DeclareGraphicsExtensions.
\caption{Comparison of $mid_{th}$ between $N=25$ vs $N=50$}
\label{midth}
\end{figure}

For moderate traffic load of $N=50$ FTP sources, changes in both average queue size and rate of change of the average queue size come into picture. Fig.~\ref{arate} shows a comparison between RED and AQMRD for $avg$ and $davg$.  In contrast to RED, it indicates that the $avg$ and $davg$ both are stabilized sufficiently by AQMRD. The $mid_{th}$ is adapted according to traffic load because it depends on the value of $davg$ that differs for different traffic loads. The parameter $mid_{th}$ is increased by one for negative value of $davg$ and decreased by one for positive value of $davg$. There is no change in $mid_{th}$ if $davg$ equals zero. The reason behind changing the $mid_{th}$ is to adjust the dropping probability subject to enhancing the performance parameters for our scheme. A negative value of $davg$ indicates that the average queue size decreases with the rate of $davg$ and a positive value of $davg$ indicates that the average queue size increases with the rate of $davg$. If the average queue size increases then it indicates that more packets are coming into the queue. To avoid larger estimates of dropped packets, AQMRD's aim is to increase the dropping probability by decreasing the $mid_{th}$ value. Similarly, for negative value of $davg$ the parameter $mid_{th}$ increases just to reduce the dropping probability.           For AQMRD gateway, we have shown in Fig.~\ref{midth} the variation of $mid_{th}$ with respect to the simulation time for $N=25$ and $N=50$. Depending on the value of $davg$, $mid_{th}$ stabilizes between $min_{th}$ and $max_{th}$. Simulation results show that $mid_{th}$ oscillates between $min_{th}=16$ and $max_{th}=48$ and stabilization varies from $min_{th}$ to $max_{th}$.
\section{Performance Evaluation}
\subsection{Simulations Under Different Scenarios}
We note in our setting that input traffic increases when the number of FTP sources reaches around 25 for a chosen delay-bandwidth product of 200 packets on the bottleneck link. This scenario is presented later in Fig.~\ref{thruN}. Therefore, we have performed the simulations for four cases $N=$ 25, 50, 75, and 100 FTP sources that respectively correspond to light traffic, moderate traffic, high traffic, and very high traffic with respect to the setting shown in Fig.~\ref{netscenario}. In our scenario, the most significant delay is the queuing delay which depends on the transmission delay. Queuing delay is proportional to the product of the number of packets transmitted and transmission delay. To have a lower queuing delay we consider high bandwidth of the bottleneck link which results in low transmission delay for a given fixed size of packets. In order to demonstrate the efficacy of the proposed algorithm, we examine the performance measures viz. throughput, link-utilization, queue size, and queuing delay through ns2 simulations.

\textbf{Scenario-1: Number of FTP sources $N=25$}

\begin{figure}[!h]
\centering
\includegraphics[angle=0,width=0.6\textwidth,height=0.5\textwidth]{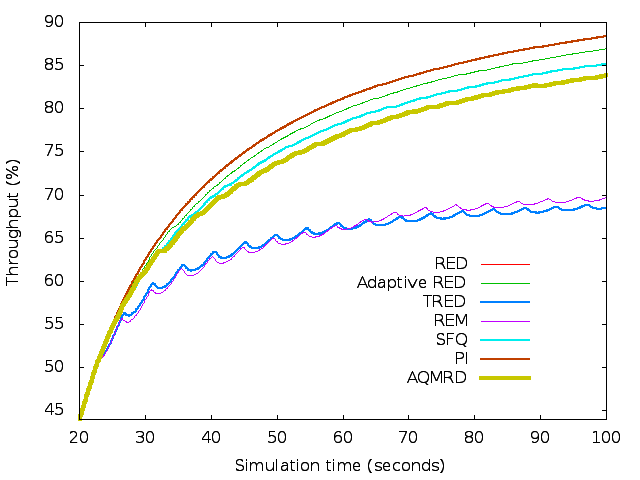}
% where an .eps filename suffix will be assumed under latex,
% and a .pdf suffix will be assumed for pdflatex; or what has been declared
% via \DeclareGraphicsExtensions.
\caption{Comparative study of throughputs for $N=25$}
\label{thru25}
\end{figure}

%We consider heavy input traffic with power law characteristics. For simulating such characteristics, we employ the algorithm due to %Feldmann and Whitt [12] who approximate a long-tail distribution by a finite mixture of exponentials. Thus superimposing outputs %from several independent sources with each source generating exponential inter-arrival time distribution, we may observe dramatic %effect upon performance of the network.
For $N=25$ FTP sources, throughput is found to increase for AQMRD when all the 25 FTP sources are superimposed but Adaptive RED is capable of achieving more throughput at low traffic loads. Fig.~\ref{thru25} shows a comparative study of throughput for different AQM algorithms. This has been achieved as a result of the adaptation of $max_p$ in the Adaptive RED scheme. Our scheme performs better than Adaptive RED for moderate traffic load as discussed in scenario-2. PI outperforms the AQM schemes because it is able to dynamically control the queue for this network scenario. This is achieved due to the fact that a sufficient buffer size is set and PI is found to be very sensitive towards buffer size.

\textbf{Scenario-2: Number of FTP sources $N=50$}

\begin{figure}[!h]
\centering
\includegraphics[angle=0,width=0.6\textwidth,height=0.5\textwidth]{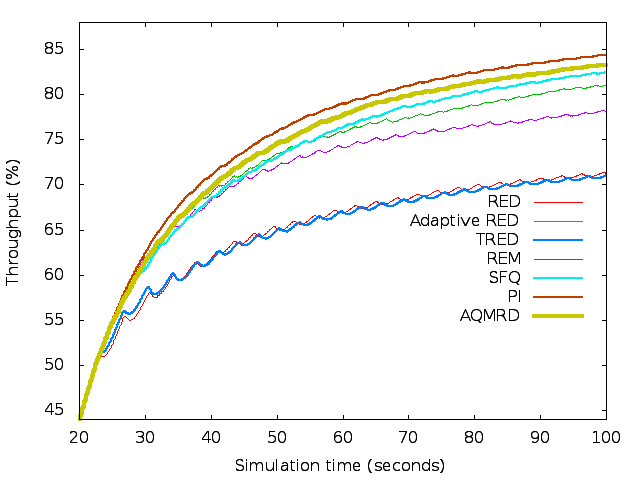}
% where an .eps filename suffix will be assumed under latex,
% and a .pdf suffix will be assumed for pdflatex; or what has been declared
% via \DeclareGraphicsExtensions.
\caption{Comparative study of throughputs for $N=50$}
\label{thru50}
\end{figure}

\begin{figure}[!h]
\centering
\includegraphics[angle=0,width=0.6 \textwidth,height=0.5 \textwidth]{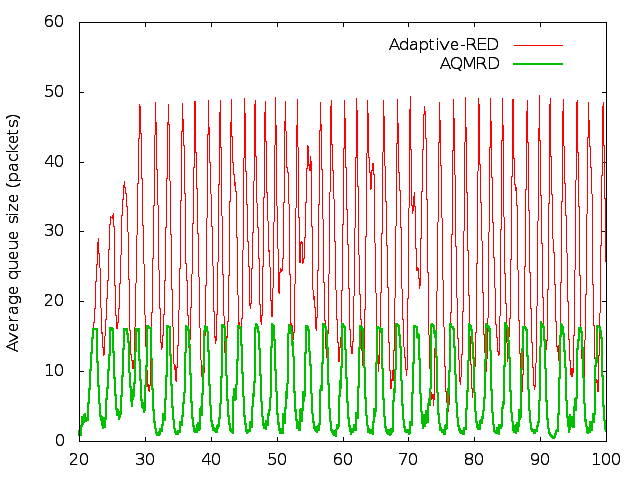}
% where an .eps filename suffix will be assumed under latex,
% and a .pdf suffix will be assumed for pdflatex; or what has been declared
% via \DeclareGraphicsExtensions.
\caption{Comparative study of $avg$ for $N=50$: Adaptive-RED vs AQMRD}
\label{avg50}
\end{figure}

Next, we increase the number of FTP sources to 50 in order to simulate a network scenario for moderate traffic load. We also compare the throughput with Adaptive RED in Fig.~\ref{thru50} which indicates that Adaptive RED fails when traffic load increases and our scheme performs better than the Adaptive RED scheme. In this setting as well for the earlier mentioned reasons, PI is seen to exhibit the best results but is closely followed by AQMRD. We compare the average queue size changes over the simulation time with Adaptive-RED, see Fig.~\ref{avg50}. The proposed AQMRD achieves better results than the other AQM algorithms except PI in terms of throughput. AQMRD also achieves better stabilization of the queue as compared to Adaptive RED. 

\textbf{Scenario-3: Number of FTP sources $N=75$}

\begin{figure}[!h]
\centering
\includegraphics[angle=0,width=0.6 \textwidth,height=0.5 \textwidth]{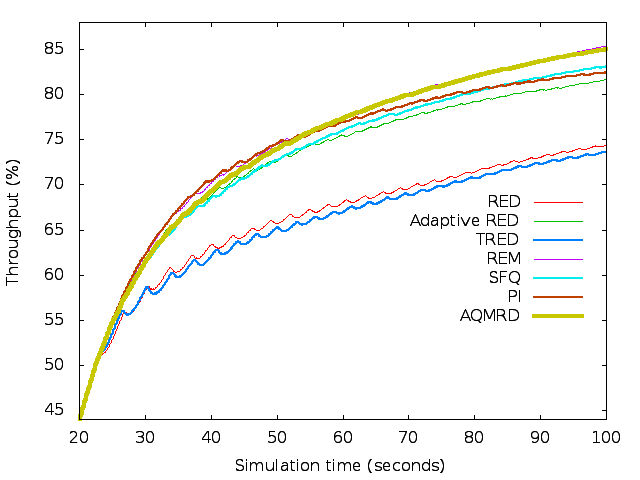}
% where an .eps filename suffix will be assumed under latex,
% and a .pdf suffix will be assumed for pdflatex; or what has been declared
% via \DeclareGraphicsExtensions.
\caption{Comparative study of throughputs for $N=75$}
\label{thru75}
\end{figure}

\begin{figure}[!h]
\centering
\includegraphics[angle=0,width=0.6 \textwidth,height=0.5 \textwidth]{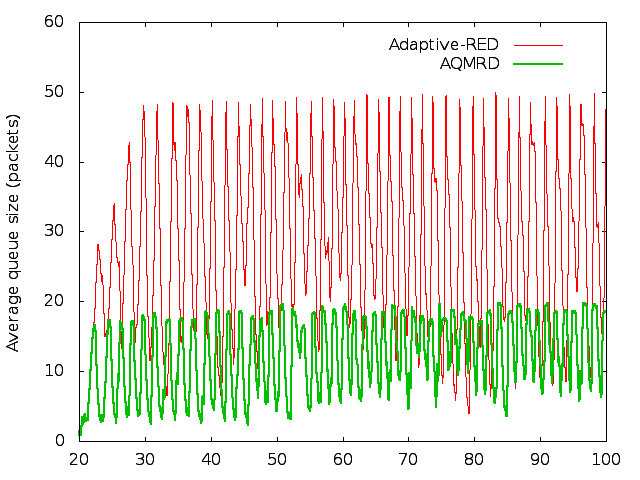}
% where an .eps filename suffix will be assumed under latex,
% and a .pdf suffix will be assumed for pdflatex; or what has been declared
% via \DeclareGraphicsExtensions.
\caption{Comparative study of $avg$ for $N=75$: Adaptive-RED vs AQMRD}
\label{avg75}
\end{figure}

In order to have a more realistic scenario, we further increase the number of FTP sources to 75. Fig.~\ref{thru75} shows the comparative study of throughput for each scheme. We observe that AQMRD shows good results here, better than PI and all the other schemes except REM. For this setting, throughput of REM is very close to our scheme. We also evaluate the average queue size for $N=75$ and observe that AQMRD is better than Adaptive RED. Fig.~\ref{avg75} compares the average queue size stabilization between Adaptive-RED and our scheme. Our scheme AQMRD gets higher stabilization for queue size than Adaptive-RED which results in lower queuing delay for our scheme. It may be noted that whereas Adaptive RED has wide fluctuations in average queue length, our algorithm AQMRD controls these significantly (see Figs.~\ref{avg50} and \ref{avg75}). Our scheme is designed so as to be able to control the traffic whether it is with low or high loads. Simulation results achieved in Fig.~\ref{thru75} show that our scheme outperforms all other schemes except REM at high traffic loads. However, the difference in performance between AQMRD and REM is marginal here.

\textbf{Scenario-4:  Number of FTP sources $N=100$}

\begin{figure}[!h]
\centering
\includegraphics[angle=0,width=0.6 \textwidth,height=0.5 \textwidth]{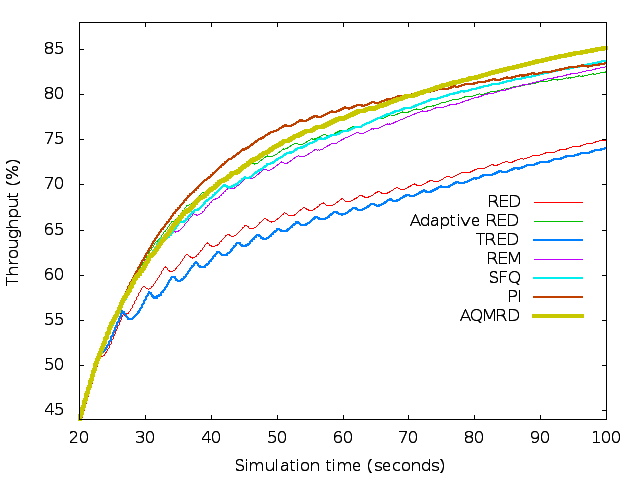}
% where an .eps filename suffix will be assumed under latex,
% and a .pdf suffix will be assumed for pdflatex; or what has been declared
% via \DeclareGraphicsExtensions.
\caption{Comparative study of throughputs for $N=100$}
\label{thru100}
\end{figure}

In order to simulate very heavy traffic, we further increase the number of FTP sources to 100. From Fig. ~\ref{thru100}, we observe that our scheme outperforms all the other AQM algorithms - RED, MRED, Adaptive-RED, TRED, REM, SFQ, and PI. Scenario-3 and scenario-4 show that our scheme is able to control the congestion better than all other schemes at high as well as very high traffic loads.

\subsection{Effect of Load}
\begin{figure}[!h]
\centering
\includegraphics[angle=0,width=0.6 \textwidth,height=0.45 \textwidth]{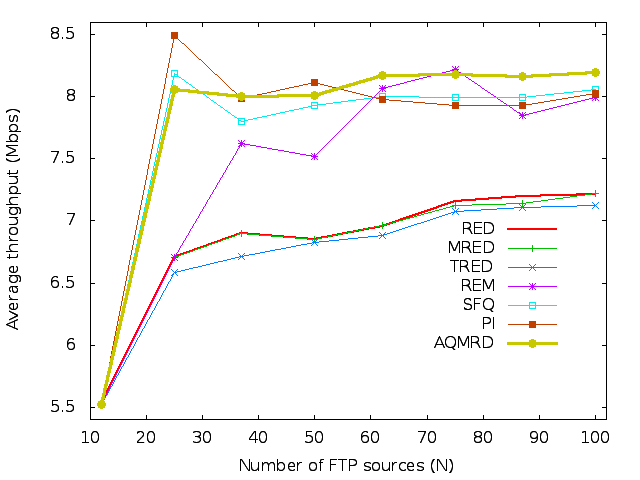}
% where an .eps filename suffix will be assumed under latex,
% and a .pdf suffix will be assumed for pdflatex; or what has been declared
% via \DeclareGraphicsExtensions.
\caption{Comparative study of the average throughput vs $N$}
\label{thruN}
\end{figure}

Next, we show the results of several experiments by using the same network parameters as before. Here, we vary the number of FTP sources from 12 to 100 and compare the throughputs in Fig. ~\ref{thruN} for eight different levels of traffic load. The purpose of these simulations on ns-2 is to evaluate the average throughput as a function of $N$. The AQMRD gateway shows improvement in throughput for each simulation and in fact shows the best results when the number of FTP sources goes beyond 54. These results show that the proposed scheme is capable of yielding good performance, regardless of the level of traffic. PI is capable of achieving the best average throughput at light traffic load but its performance degrades at high traffic loads. In comparison, our scheme is competitive against PI and in fact, consistently outperforms PI at traffic loads beyond 54 sources. The REM algorithm shows good results in a narrow change and is marginally better than AQMRD for $N=75$ but does not compare favourably at other load levels. Unlike the other algorithms, our scheme exercises better control over the network traffic.
\subsection{Effect of $max_{th}$}

\begin{figure}[!h]
\centering
\includegraphics[angle=0,width=0.6 \textwidth,height=0.45\textwidth]{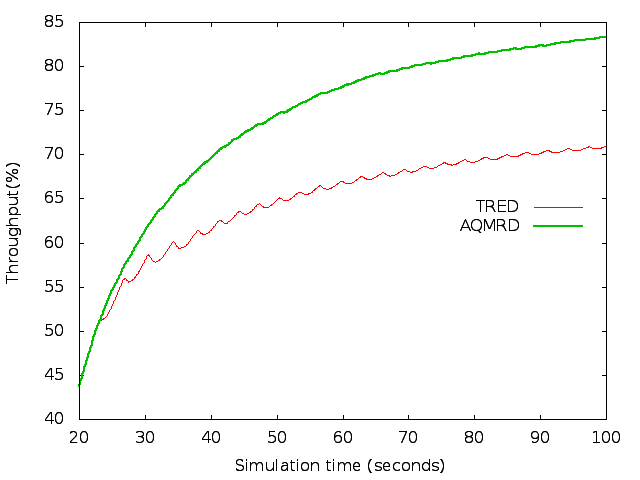}
% where an .eps filename suffix will be assumed under latex,
% and a .pdf suffix will be assumed for pdflatex; or what has been declared
% via \DeclareGraphicsExtensions.
\caption{Throughput comparison of AQMRD vs TRED, $ max_{th}=48$}
\label{thrua}
\end{figure}

\begin{figure}[!h]
\centering
\includegraphics[angle=0,width=0.6 \textwidth,height=0.45\textwidth]{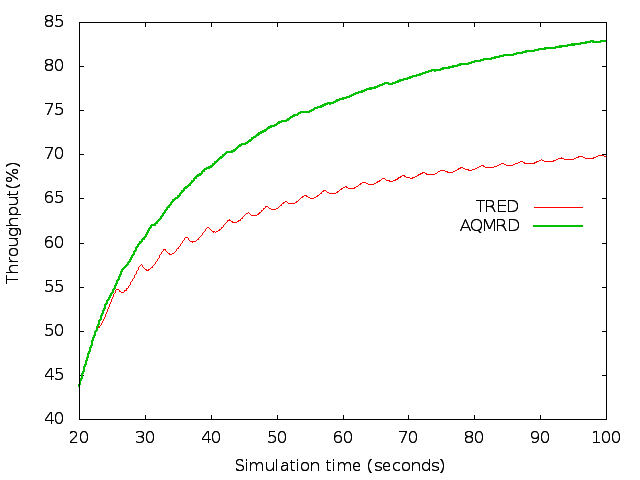}
% where an .eps filename suffix will be assumed under latex,
% and a .pdf suffix will be assumed for pdflatex; or what has been declared
% via \DeclareGraphicsExtensions.
\caption{Throughput comparison of AQMRD vs TRED, $max_{th}=30$}
\label{thrub}
\end{figure}

\begin{figure}[!h]
\centering
\includegraphics[angle=0,width=0.6 \textwidth,height=0.45\textwidth]{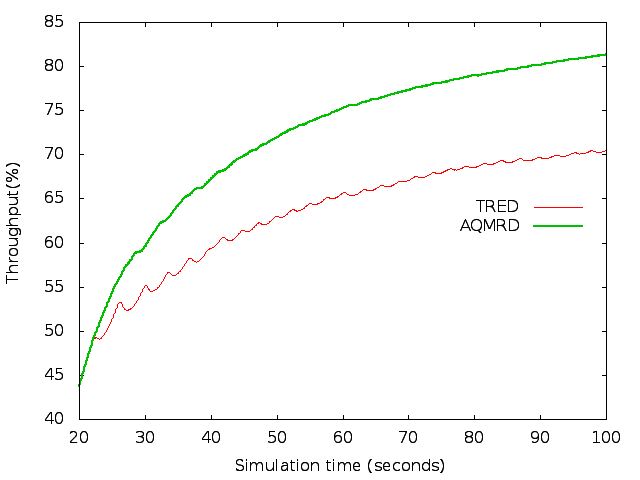}
% where an .eps filename suffix will be assumed under latex,
% and a .pdf suffix will be assumed for pdflatex; or what has been declared
% via \DeclareGraphicsExtensions.
\caption{Throughput comparison of AQMRD vs TRED, $ max_{th}=18$}
\label{thruc}
\end{figure}

\begin{figure}[!h]
\centering
\includegraphics[angle=0,width=0.6\textwidth,height=0.45\textwidth]{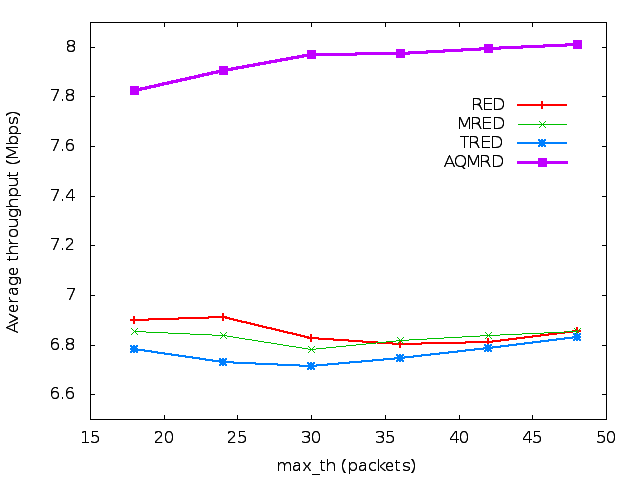}
% where an .eps filename suffix will be assumed under latex,
% and a .pdf suffix will be assumed for pdflatex; or what has been declared
% via \DeclareGraphicsExtensions.
\caption{Comparative study of the average throughput vs $max_{th}$: RED, MRED, TRED and AQMRD}
\label{thruvsmax}
\end{figure}

Next, we consider the same network parameters as before and evaluate the link utilization or relative throughput (i.e., throughput / bandwidth) for three different levels of $max_{th}$. We compare the relative throughput of our scheme with that of the recently proposed scheme TRED. It is remarkable that the relative throughput for our scheme approximately increases between 12-16 \% over TRED for three different levels of $max_{th}$, see Figs.~\ref{thrua} - \ref{thruc}. We have not shown the relative throughput for RED and MRED because there is only a marginal change in throughput in the schemes over TRED. In Fig.~\ref{thruvsmax} we compare the average throughput of AQMRD with RED, MRED and TRED. The average throughput is computed as the ratio of total number of bytes sent to the total simulation time. The average throughput in each case is evaluated for six different simulations as $max_{th}$ varies from 18 to 48 in equal intervals. All the parameters and the variables other than $max_{th}$ are the same as before. The performance comparisons of the average throughput versus $max_{th}$ of our algorithm with RED, MRED, and TRED are given in Fig.~\ref{thruvsmax}. We evaluate the average and relative throughput for different levels of $max_{th}$ and observe from Figs.~\ref{thrua}~-~\ref{thruvsmax} that AQMRD performs better than RED, MRED, and TRED, in the relative throughput metric as well. 
 

\subsection{Effect of Buffer Size}
\begin{figure}[!h]
\centering
\includegraphics[angle=0,width=0.6\textwidth,height=0.45\textwidth]{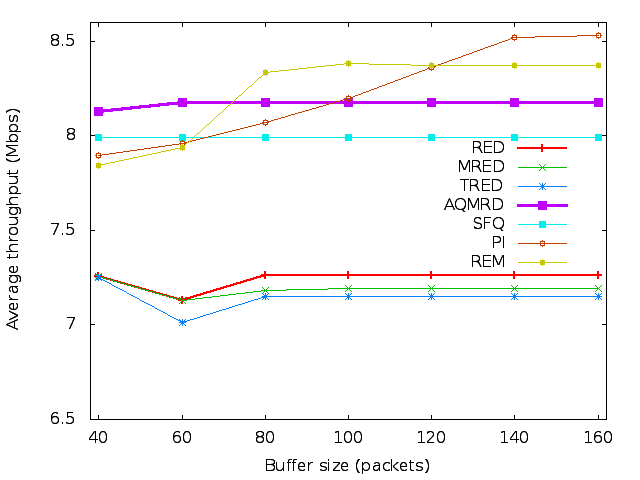}
% where an .eps filename suffix will be assumed under latex,
% and a .pdf suffix will be assumed for pdflatex; or what has been declared
% via \DeclareGraphicsExtensions.
\caption{Comparative study of the average throughput vs buffer size: RED, MRED, TRED, SFQ, REM, PI and AQMRD}
\label{thruvsB}
\end{figure}

In order to check the sensitivity of the buffer size in high traffic load, we performed simulations with varying  buffer sizes between $40$ to $160$ and fixing $max_{th}$ to 48 for each simulation. We consider the simulation for high trafic load with $N=75$ FTP sources. We have performed seven simulations to carry out the results of throughput as a function of the buffer size. Fig.~\ref{thruvsB} indicates that our proposed scheme uniformly achieves more than 12  \% higher throughput than the other algorithms except the two schemes REM and PI. Simulation results also show that PI and REM are highly sensitive to the buffer size. AQMRD outperforms all AQM schemes at low buffer size due to the introduction of the rate of change of average queue size parameter. In fact, our scheme is more robust and shows the least sensitivity or variation in performance as a function of the buffer size. 
\subsection{Queuing Delay}
\begin{figure}[!h]
\centering
\includegraphics[angle=0,width=0.6\textwidth,height=0.45\textwidth]{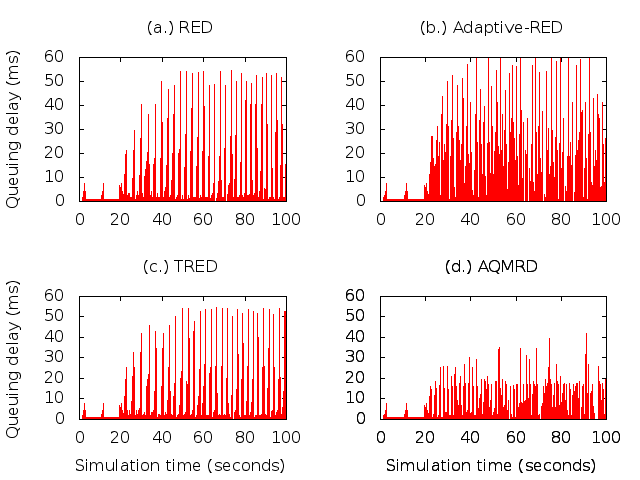}
\includegraphics[angle=0,width=0.6\textwidth,height=0.45\textwidth]{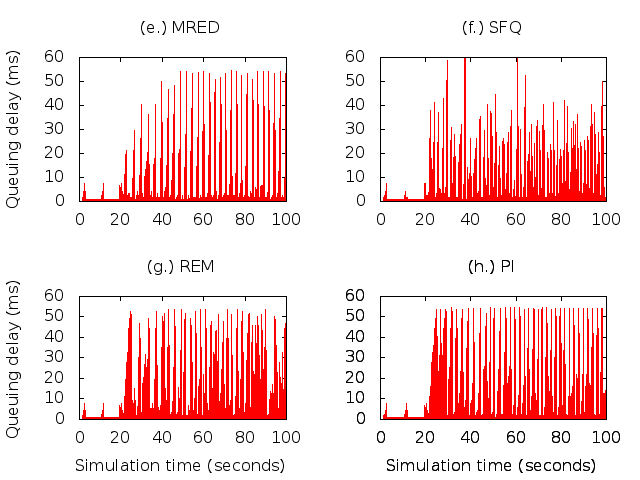}
% where an .eps filename suffix will be assumed under latex,
% and a .pdf suffix will be assumed for pdflatex; or what has been declared
% via \DeclareGraphicsExtensions.
\caption{Comparisons of queuing delay for $N=75$}
\label{del75_2}
\end{figure}

\begin{figure}[!h]
\centering
\includegraphics[angle=0,width=0.6\textwidth,height=0.45\textwidth]{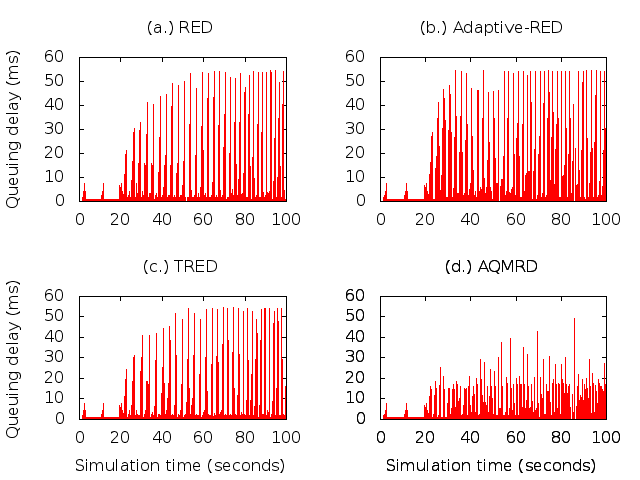}
\includegraphics[angle=0,width=0.6\textwidth,height=0.45\textwidth]{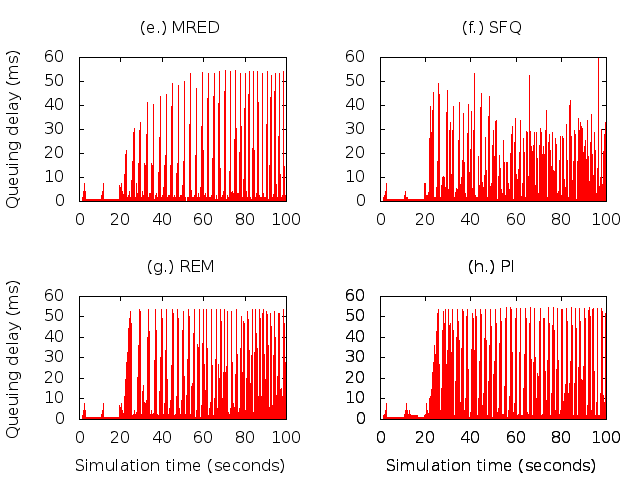}
% where an .eps filename suffix will be assumed under latex,
% and a .pdf suffix will be assumed for pdflatex; or what has been declared
% via \DeclareGraphicsExtensions.
\caption{Comparisons of queuing delay for $N=100$}
\label{del100_2}
\end{figure}

 Queuing delay plays an important role under high traffic condition. We have performed simulation experiments to compare the queuing delay and throughput for high traffic with $N=75$ FTP sources. TRED achieves a low delay with same throughput as RED under high load. Our scheme reduces the queuing delay significantly over TRED even while resulting in higher throughput.  Fig.~\ref{del75_2} shows that for $N=75$, AQMRD exhibits smaller queuing delay in relation to RED, Adaptive-RED, and TRED. The AQMRD gateway achieves significantly lower queuing delay and higher throughput than each of the schemes RED, MRED, TRED, and Adaptive-RED. From Fig.~\ref{del100_2}, we observed that there is much improvement  in queuing delay when using our scheme AQMRD. We also obtain good throughput as shown in Fig.~\ref{thruvsB} which conclusively demonstrates that AQMRD achieves low delay without affecting the goal of achieving high throughput. 
\subsection{Mean Values of Performance Measures}
Expected values of performance measures such as queuing delay, average queue size, instantaneous queue size, and average loss-ratio are given in Tables~\ref{tab1},~\ref{tab3},~\ref{tab5}, and ~\ref{tab7}, respectively. PI is unable to attain the acceptable queuing delay as its expected queuing delay increases by 300\% drastically with respect to RED. On the other hand, it can be seen that the AQMRD gateway achieves upto 17.74\% reduction in expected queuing delay. Tables~\ref{tab2},~\ref{tab4}, and ~\ref{tab6} show the percentage reduction in queuing delay, expected value of the average queue size, and expected value of instantaneous queue size each with respect to the RED scheme. Table~\ref{tab8} shows the percentage increase in loss-ratio for each scheme with respect to RED. This reduction in queuing delay is on account of lower expected value of average and instantaneous queue sizes observed in AQMRD. Feng et al. (see section II. C of~\cite{sp21}) pointed out that it may not be possible to achieve both high link utilization / throughput and low queuing delay simultaneously. Our results, however, demonstrate that this can be done, i.e., achieving high throughput and low queuing delay albeit with a higher packet loss rate. It should be emphasized that our proposed algorithm significantly improves both the QoS parameters.
\begin{table}[!h]
\begin{center}
\caption{\textsc{Comparative Performance Results: Queuing Delay}}
\label{tab1}
\begin{tabular}{|l|r|r|r|r|r|r|r|}
\hline
& \multicolumn{5}{p{5.3 cm}|}%
{\centering \textbf{E[Queuing Delay] (ms)}}\\
\cline{2-5}
\multicolumn{1}{|c|}{\centering \textbf{\textit{AQM Algorithms}}}
& \multicolumn{1}{c|}{$N=25$}
& \multicolumn{1}{c|}{$N=50$}
& \multicolumn{1}{c|}{$N=75$}
& \multicolumn{1}{c|}{$N=100$}\\
\cline{1-5}
\multicolumn{1}{|c|}{RED}&\multicolumn{1}{|c|}{0.03071} & \multicolumn{1}{c|}{0.02671}& \multicolumn{1}{c|}{0.02656}
& \multicolumn{1}{c|}{0.02571}\\

\cline{1-5}
\multicolumn{1}{|c|}{MRED}&\multicolumn{1}{|c|}{0.03071} & \multicolumn{1}{c|}{0.02672}& \multicolumn{1}{c|}{0.02807}
& \multicolumn{1}{c|}{0.02576}\\

\cline{1-5}
\multicolumn{1}{|c|}{Adaptive-RED}&\multicolumn{1}{|c|}{0.06678} & \multicolumn{1}{c|}{0.04092}& \multicolumn{1}{c|}{0.03209}
& \multicolumn{1}{c|}{0.03208}\\

\cline{1-5}
\multicolumn{1}{|c|}{SFQ}&\multicolumn{1}{|c|}{0.03651} & \multicolumn{1}{c|}{0.02206}& \multicolumn{1}{c|}{0.02086}
& \multicolumn{1}{c|}{0.02085}\\

\cline{1-5}
\multicolumn{1}{|c|}{REM}&\multicolumn{1}{|c|}{0.03071} & \multicolumn{1}{c|}{0.03802}& \multicolumn{1}{c|}{0.04191}
& \multicolumn{1}{c|}{0.03425}\\

\cline{1-5}
\multicolumn{1}{|c|}{PI}&\multicolumn{1}{|c|}{0.09313} & \multicolumn{1}{c|}{0.05348}& \multicolumn{1}{c|}{0.04498}
& \multicolumn{1}{c|}{0.06083}\\

\cline{1-5}
\multicolumn{1}{|c|}{TRED}&\multicolumn{1}{|c|}{0.03437} & \multicolumn{1}{c|}{0.02958}& \multicolumn{1}{c|}{0.02642}
& \multicolumn{1}{c|}{0.02690}\\

\cline{1-5}
\multicolumn{1}{|c|}{AQMRD}&\multicolumn{1}{|c|}{0.02871} & \multicolumn{1}{c|}{0.02413}& \multicolumn{1}{c|}{0.02259}
& \multicolumn{1}{c|}{0.02115}\\
\hline
\end{tabular}
\end{center}
\end{table}

\begin{table}[!h]
\begin{center}
\caption{\textsc{Percentage Reduction in Expected Queuing Delay with respect to RED}}
\label{tab2}
\begin{tabular}{|l|r|r|r|r|r|r|r|}
\hline
& \multicolumn{5}{p{6.1 cm}|}%
{\centering \textbf{Number of FTP Sources $(N)$}}\\
\cline{2-5}
\multicolumn{1}{|c|}{\centering \textbf{\textit{AQM Algorithms}}}
& \multicolumn{1}{c|}{$N=25$}
& \multicolumn{1}{c|}{$N=50$}
& \multicolumn{1}{c|}{$N=75$}
& \multicolumn{1}{c|}{$N=100$}\\
\cline{1-5}
\multicolumn{1}{|c|}{MRED }&\multicolumn{1}{|c|}{0\%} & \multicolumn{1}{c|}{$-$0.04\%}& \multicolumn{1}{c|}{$-$5.69\%} & \multicolumn{1}{c|}{$-$0.19\%}\\

\cline{1-5}
\multicolumn{1}{|c|}{Adaptive-RED}&\multicolumn{1}{|c|}{$-$117.45\%} & \multicolumn{1}{c|}{$-$53.20\%}& \multicolumn{1}{c|}{$-$20.82\%} & \multicolumn{1}{c|}{$-$0.19\%}\\

 \cline{1-5}
\multicolumn{1}{|c|}{SFQ}&\multicolumn{1}{|c|}{$-$18.89\%} & \multicolumn{1}{c|}{$+$17.41\%}& \multicolumn{1}{c|}{$+$21.46\%} & \multicolumn{1}{c|}{$+$18.90\%}\\

\cline{1-5}
\multicolumn{1}{|c|}{REM}&\multicolumn{1}{|c|}{0\%} & \multicolumn{1}{c|}{$-$42.34\%}& \multicolumn{1}{c|}{$+$57.79\%} & \multicolumn{1}{c|}{$-$33.22\%}\\

\cline{1-5}
\multicolumn{1}{|c|}{PI}&\multicolumn{1}{|c|}{$-$300.25\%} & \multicolumn{1}{c|}{$-$100.22\%}& \multicolumn{1}{c|}{$-$69.35\%} & \multicolumn{1}{c|}{$-$136.60\%}\\

\cline{1-5}
\multicolumn{1}{|c|}{TRED }&\multicolumn{1}{|c|}{$-$11.91\%} & \multicolumn{1}{c|}{$-$10.75\%}& \multicolumn{1}{c|}{$+$0.53\%} & \multicolumn{1}{c|}{$-$4.63\%}\\

\cline{1-5}
\multicolumn{1}{|c|}{AQMRD}&\multicolumn{1}{|c|}{$+$6.51\%} & \multicolumn{1}{c|}{$+$9.66\%}& \multicolumn{1}{c|}{$+$14.95\%} & \multicolumn{1}{c|}{$+$17.74\%}\\
\hline
\end{tabular}
\end{center}
\end{table}

\begin{table}[!h]
\begin{center}
\caption{\textsc{Comparative Performance Results: Time Average of Average Queue Size}}
\label{tab3}
\begin{tabular}{|l|r|r|r|r|r|r|r|}
\hline
& \multicolumn{5}{p{5.3 cm}|}%
{\centering \textbf{E[Average Queue Size] (packets)}}\\
\cline{2-5}
\multicolumn{1}{|c|}{\centering \textbf{\textit{AQM Algorithms}}}
& \multicolumn{1}{c|}{$N=25$}
& \multicolumn{1}{c|}{$N=50$}
& \multicolumn{1}{c|}{$N=75$}
& \multicolumn{1}{c|}{$N=100$}\\
\cline{1-5}
\multicolumn{1}{|c|}{RED}&\multicolumn{1}{|c|}{6.41} & \multicolumn{1}{c|}{11.51}& \multicolumn{1}{c|}{14.02}
& \multicolumn{1}{c|}{14.31}\\

\cline{1-5}
\multicolumn{1}{|c|}{MRED}&\multicolumn{1}{|c|}{6.41} & \multicolumn{1}{c|}{11.60}& \multicolumn{1}{c|}{14.18}
& \multicolumn{1}{c|}{14.51}\\

\cline{1-5}
\multicolumn{1}{|c|}{Adaptive-RED}&\multicolumn{1}{|c|}{21.12} & \multicolumn{1}{c|}{23.15}& \multicolumn{1}{c|}{24.67}
& \multicolumn{1}{c|}{21.01}\\

\cline{1-5}
\multicolumn{1}{|c|}{TRED}&\multicolumn{1}{|c|}{6.97} & \multicolumn{1}{c|}{11.54}& \multicolumn{1}{c|}{14.01}
& \multicolumn{1}{c|}{14.80}\\

\cline{1-5}
\multicolumn{1}{|c|}{AQMRD}&\multicolumn{1}{|c|}{7.06} & \multicolumn{1}{c|}{7.09}& \multicolumn{1}{c|}{11.30}
& \multicolumn{1}{c|}{11.53}\\
\hline
\end{tabular}
\end{center}
\end{table}

\begin{table}[!h]
\begin{center}
\caption{\textsc{Percentage Reduction in the Expected Average Queue Size with respect to RED}}
\label{tab4}
\begin{tabular}{|l|r|r|r|r|r|r|r|}
\hline
& \multicolumn{5}{p{5.95 cm}|}%
{\centering \textbf{Number of FTP Sources $(N)$}}\\
\cline{2-5}
\multicolumn{1}{|c|}{\centering \textbf{\textit{AQM Algorithms}}}
& \multicolumn{1}{c|}{$N=25$}
& \multicolumn{1}{c|}{$N=50$}
& \multicolumn{1}{c|}{$N=75$}
& \multicolumn{1}{c|}{$N=100$}\\
\cline{1-5}
\multicolumn{1}{|c|}{MRED}&\multicolumn{1}{|c|}{0\%} & \multicolumn{1}{c|}{$-$0.78\%}& \multicolumn{1}{c|}{$-$1.14\%} & \multicolumn{1}{c|}{$-$1.40\%}\\
\cline{1-5}
\multicolumn{1}{|c|}{Adaptive-RED}&\multicolumn{1}{|c|}{$-$229.49\%} & \multicolumn{1}{c|}{$-$101.13\%}& \multicolumn{1}{c|}{$-$75.96\%} & \multicolumn{1}{c|}{$-$46.82\%}\\
\cline{1-5}
\multicolumn{1}{|c|}{TRED}&\multicolumn{1}{|c|}{$-$8.74\%} & \multicolumn{1}{c|}{$-$0.26\%}& \multicolumn{1}{c|}{$+$0.071\%} & \multicolumn{1}{c|}{$-$3.42\%}\\
\cline{1-5}
\multicolumn{1}{|c|}{AQMRD }&\multicolumn{1}{|c|}{$-$10.14\%} & \multicolumn{1}{c|}{$+$38.40\%}& \multicolumn{1}{c|}{$+$19.40\%} & \multicolumn{1}{c|}{$+$19.43\%}\\
\hline
\end{tabular}
\end{center}
\end{table}

\begin{table}[!h]
\begin{center}
\caption{\textsc{Comparative Performance Results: Expected Instantaneous Queue Size}}
\label{tab5}
\begin{tabular}{|l|r|r|r|r|r|r|r|}
\hline
& \multicolumn{5}{p{5.3 cm}|}%
{\centering \textbf{E[Instantaneous Queue Size] (packets)}}\\
\cline{2-5}
\multicolumn{1}{|c|}{\centering \textbf{\textit{AQM Algorithms}}}
& \multicolumn{1}{c|}{$N=25$}
& \multicolumn{1}{c|}{$N=50$}
& \multicolumn{1}{c|}{$N=75$}
& \multicolumn{1}{c|}{$N=100$}\\
\cline{1-5}
\multicolumn{1}{|c|}{RED}&\multicolumn{1}{|c|}{6.60} & \multicolumn{1}{c|}{13.23}& \multicolumn{1}{c|}{17.83}
& \multicolumn{1}{c|}{19.34}\\

\cline{1-5}
\multicolumn{1}{|c|}{MRED}&\multicolumn{1}{|c|}{6.60} & \multicolumn{1}{c|}{13.23}& \multicolumn{1}{c|}{18.94}
& \multicolumn{1}{c|}{19.65}\\

\cline{1-5}
\multicolumn{1}{|c|}{Adaptive-RED}&\multicolumn{1}{|c|}{19.11} & \multicolumn{1}{c|}{26.32}& \multicolumn{1}{c|}{28.67}
& \multicolumn{1}{c|}{28.01}\\

\cline{1-5}
\multicolumn{1}{|c|}{TRED}&\multicolumn{1}{|c|}{7.14} & \multicolumn{1}{c|}{13.11}& \multicolumn{1}{c|}{17.76}
& \multicolumn{1}{c|}{20.36}\\

\cline{1-5}
\multicolumn{1}{|c|}{AQMRD}&\multicolumn{1}{|c|}{7.91} & \multicolumn{1}{c|}{13.39}& \multicolumn{1}{c|}{15.27}
& \multicolumn{1}{c|}{15.56}\\
\hline
\end{tabular}
\end{center}
\end{table}

\begin{table}[!h]
\begin{center}
\caption{\textsc{Percentage Reduction in the Expected Instantaneous Queue Size with respect to RED}}
\label{tab6}
\begin{tabular}{|l|r|r|r|r|r|r|r|}
\hline
& \multicolumn{5}{p{5.85 cm}|}%
{\centering \textbf{Number of FTP Sources $(N)$}}\\
\cline{2-5}
\multicolumn{1}{|c|}{\centering \textbf{\textit{AQM Algorithms}}}
& \multicolumn{1}{c|}{$N=25$}
& \multicolumn{1}{c|}{$N=50$}
& \multicolumn{1}{c|}{$N=75$}
& \multicolumn{1}{c|}{$N=100$}\\
\cline{1-5}
\multicolumn{1}{|c|}{MRED }&\multicolumn{1}{|c|}{0\%} & \multicolumn{1}{c|}{0\%}& \multicolumn{1}{c|}{$-$6.23\%} & \multicolumn{1}{c|}{$-$1.60\%}\\
\cline{1-5}
\multicolumn{1}{|c|}{Adaptive-RED}&\multicolumn{1}{|c|}{$-$189.55\%} & \multicolumn{1}{c|}{$-$98.94\%}& \multicolumn{1}{c|}{$-$60.80\%} & \multicolumn{1}{c|}{$-$44.83\%}\\
\cline{1-5}
\multicolumn{1}{|c|}{TRED }&\multicolumn{1}{|c|}{$-$8.18\%} & \multicolumn{1}{c|}{$+$0.91\%}& \multicolumn{1}{c|}{$+$0.40\%} & \multicolumn{1}{c|}{$-$5.27\%}\\
\cline{1-5}
\multicolumn{1}{|c|}{AQMRD}&\multicolumn{1}{|c|}{$-$19.85\%} & \multicolumn{1}{c|}{$-$1.21\%}& \multicolumn{1}{c|}{$+$14.36\%} & \multicolumn{1}{c|}{$+$19.55\%}\\
\hline
\end{tabular}
\end{center}
\end{table}

\begin{table}[!h]
\begin{center}
\caption{\textsc{Comparative Performance Results: Average Loss Ratio}}
\label{tab7}
\begin{tabular}{|l|r|r|r|r|r|r|r|}
\hline
& \multicolumn{5}{p{5.25 cm}|}%
{\centering \textbf{Average Loss-ratio}}\\
\cline{2-5}
\multicolumn{1}{|c|}{\centering \textbf{\textit{AQM Algorithms}}}
& \multicolumn{1}{c|}{$N=25$}
& \multicolumn{1}{c|}{$N=50$}
& \multicolumn{1}{c|}{$N=75$}
& \multicolumn{1}{c|}{$N=100$}\\
\cline{1-5}
\multicolumn{1}{|c|}{RED}&\multicolumn{1}{|c|}{0.578} & \multicolumn{1}{c|}{1.771}& \multicolumn{1}{c|}{2.690}
& \multicolumn{1}{c|}{3.067}\\

\cline{1-5}
\multicolumn{1}{|c|}{MRED}&\multicolumn{1}{|c|}{0.578} & \multicolumn{1}{c|}{1.771}& \multicolumn{1}{c|}{2.911}
& \multicolumn{1}{c|}{3.012}\\

\cline{1-5}
\multicolumn{1}{|c|}{Adaptive-RED}&\multicolumn{1}{|c|}{0.200} & \multicolumn{1}{c|}{1.94}& \multicolumn{1}{c|}{2.833}
& \multicolumn{1}{c|}{3.078}\\

\cline{1-5}
\multicolumn{1}{|c|}{TRED}&\multicolumn{1}{|c|}{0.666} & \multicolumn{1}{c|}{1.819}& \multicolumn{1}{c|}{2.910}
& \multicolumn{1}{c|}{3.314}\\

\cline{1-5}
\multicolumn{1}{|c|}{REM}&\multicolumn{1}{|c|}{0.578} & \multicolumn{1}{c|}{1.262}& \multicolumn{1}{c|}{1.819}
& \multicolumn{1}{c|}{2.242}\\

\cline{1-5}
\multicolumn{1}{|c|}{SFQ}&\multicolumn{1}{|c|}{0.374} & \multicolumn{1}{c|}{1.845}& \multicolumn{1}{c|}{2.790}
& \multicolumn{1}{c|}{3.077}\\

\cline{1-5}
\multicolumn{1}{|c|}{PI}&\multicolumn{1}{|c|}{0.199} & \multicolumn{1}{c|}{1.249}& \multicolumn{1}{c|}{1.908}
& \multicolumn{1}{c|}{2.177}\\

\cline{1-5}
\multicolumn{1}{|c|}{AQMRD}&\multicolumn{1}{|c|}{0.638} & \multicolumn{1}{c|}{2.771}& \multicolumn{1}{c|}{4.086}
& \multicolumn{1}{c|}{4.490}\\

\hline
\end{tabular}
\end{center}
\end{table}

\begin{table}[!h]
\begin{center}
\caption{\textsc{Percentage Increase in Average Loss-ratio over RED}}
\label{tab8}
\begin{tabular}{|l|r|r|r|r|r|r|r|}
\hline
& \multicolumn{5}{p{5.7 cm}|}%
{\centering \textbf{Number of FTP Sources $(N)$}}\\
\cline{2-5}
\multicolumn{1}{|c|}{\centering \textbf{\textit{AQM Algorithms}}}
& \multicolumn{1}{c|}{$N=25$}
& \multicolumn{1}{c|}{$N=50$}
& \multicolumn{1}{c|}{$N=75$}
& \multicolumn{1}{c|}{$N=100$}\\
\cline{1-5}
\multicolumn{1}{|c|}{MRED}&\multicolumn{1}{|c|}{0\%} & \multicolumn{1}{c|}{0\%}& \multicolumn{1}{c|}{$+$8.22\%} & \multicolumn{1}{c|}{\centering\textbf{$-$}1.79\%}\\
\cline{1-5}
\multicolumn{1}{|c|}{Adaptive-RED}&\multicolumn{1}{|c|}{$-$65.40\%} & \multicolumn{1}{c|}{$+$9.54\%}& \multicolumn{1}{c|}{$+$5.32\%} & \multicolumn{1}{c|}{$+$0.36\%}\\
\cline{1-5}
\multicolumn{1}{|c|}{TRED }&\multicolumn{1}{|c|}{$+$15.19\%} & \multicolumn{1}{c|}{$+$2.71\%}& \multicolumn{1}{c|}{$+$8.18\%} & \multicolumn{1}{c|}{$+$8.05\%}\\
\cline{1-5}

\multicolumn{1}{|c|}{REM}&\multicolumn{1}{|c|}{0\%} & \multicolumn{1}{c|}{$+$28.74\%}& \multicolumn{1}{c|}{$+$32.78\%} & \multicolumn{1}{c|}{$+$26.89\%}\\
\cline{1-5}

\multicolumn{1}{|c|}{SFQ}&\multicolumn{1}{|c|}{$+$35.29\%} & \multicolumn{1}{c|}{$-$4.18\%}& \multicolumn{1}{c|}{$-$3.72\%} & \multicolumn{1}{c|}{$-$0.33\%}\\
\cline{1-5}

\multicolumn{1}{|c|}{PI}&\multicolumn{1}{|c|}{$+$65.57\%} & \multicolumn{1}{c|}{$+$29.48\%}& \multicolumn{1}{c|}{$+$29.07\%} & \multicolumn{1}{c|}{$+$29.02\%}\\
\cline{1-5}

\multicolumn{1}{|c|}{AQMRD }&\multicolumn{1}{|c|}{$+$10.38\%} & \multicolumn{1}{c|}{$+$56.47\%}& \multicolumn{1}{c|}{$+$51.90\%} & \multicolumn{1}{c|}{$+$46.40\%}\\
\hline
\end{tabular}
\end{center}
\end{table}

\section{Conclusions}
We presented in this paper a new AQMRD algorithm that incorporates both average queue size and its rate of change, and is found to achieve significantly better performance than RED as well as its variants in the presence of non-stationary heavy traffic. An important aspect of the proposed approach is that it incorporates the rate of change in queue size as an additional parameter. AQMRD prevents the frequent crossing of maximum threshold by the average queue size and rapidly responds to congestion before the overflow of packets occurs. As noted by Feng et al.~\cite{sp8}, the principal goal of the congestion controlling algorithm is to maintain low level of the queue size so as to keep low delay. Based on extensive numerical experiments, we have found that AQMRD has outperformed the existing AQM algorithms RED, MRED, TRED and Adaptive-RED algorithms as can be seen in Figs.~\ref{thru50} - \ref{del100_2}. Our scheme AQMRD also outperforms the PI and REM at low buffer sizes as also high traffic loads. The proposed AQMRD algorithm has the advantage of reducing the delay and queue size in comparison to the existing RED algorithm but PI is seen to have the worst queuing delay. An important finding here is that the AQMRD gateway reduces the expected value of average queue size by 38.4\% for 50 FTP sources. We may point out that the inclusion of the new parameters corresponding to the rate of change of average queue size as well as the $mid_{th}$ threshold level have played a significant role in enhancing the performance of the algorithm. An area of future enquiry would be to optimize the values of $mid_{th}$ and the other parameters $max_{th}, min_{th}, w_q, max_p$ etc. using a stochastic optimization approach.

% or


%\section*{Acknowledgment}

%The authors would like to thank the reviewer.

\ifCLASSOPTIONcaptionsoff
  \newpage
\fi

\end{document}